# LEVERAGING SECURITY OBSERVABILITY TO STRENGTHEN SECURITY OF DIGITAL ECOSYSTEM ARCHITECTURE

Renjith Ramachandran

Independent Researcher and Solutions Architect
renjith_r02@yahoo.com

## ABSTRACT

*In the current fast-paced digital environment, enterprises are striving to offer a seamless and integrated customer experience across multiple touchpoints. This improved experience often leads to higher conversion rates and increased customer loyalty. To deliver such an experience, enterprises must think beyond the traditional boundaries of their architecture. The architecture of the digital ecosystem is expanding and becoming more complex, achieved either by developing advanced features in-house or by integrating with third-party solutions, thus extending the boundaries of the enterprise architecture.*

*This complexity poses significant challenges for both observability and security in a digital ecosystem, both of which are essential for maintaining robust and resilient systems. Observability entails monitoring and understanding the internal state of a system through logging, tracing, and metrics collection, allowing organizations to diagnose performance issues and detect anomalies in real time. Meanwhile, security is focused on protecting sensitive data and ensuring service integrity by defending against threats and vulnerabilities. The data collected through these observability practices can be analyzed to identify patterns and detect potential security threats or data leaks.*

*This paper examines the interconnections between observability and security within digital ecosystem architectures, emphasizing how improved observability can strengthen security measures. It discusses the additional data that observability practices can capture across different application layers and how this data can be leveraged to enhance overall system security. The paper also discusses studies conducted in the AI/ML field aimed at enhancing security through the use of observability. These studies explore how advanced machine learning techniques can be applied to observability data to improve security measures and detect anomalies more effectively.*

## 1. INTRODUCTION

Enterprise architecture is evolving with the rise of digital initiatives, transitioning to an ecosystem architecture that includes multiple third-party systems. In this digital era, cybersecurity is one of the most critical concerns for enterprises, as it protects organizations from a variety of threats that arise from their growing reliance on technology [1]. As digital systems become more integrated into everyday operations, the risks associated with cyberattacks, data breaches, and privacy violations increase. These security incidents often result in significant financial losses as well [2].

Building a secure system requires implementing security measures across every layer of the application, from the developer's local environment and secure coding practices to infrastructure security. As the application grows and becomes more complex, maintaining security becomes increasingly challenging. The "shift-left" approach helps address this by incorporating security earlier in the development process. This typically involves steps such as

Static Application Security Testing (SAST), which detects vulnerabilities in the code before it is deployed in any environment.

Once an application is live, enterprises typically use Dynamic Application Security Testing (DAST) scans for penetration testing. However, DAST scans have limitations in identifying all the types of the attacks as its more human generated tests [7]. A more effective approach is required to identify threats in real-time and take appropriate actions. To address this, a mechanism is needed to collect real-time data from various systems, analyze it, and use it for threat detection. This paper demonstrates how data gathered by observability tools from different layers of an application can be leveraged for threat detection and remediation.

## 2. OBSERVABILITY

Observability is the process of gaining insight into the state of a system by analyzing logs, metrics, and traces collected from it. Unlike monitoring, observability focuses on providing the necessary data to understand the "why" and "how" behind system behavior, allowing for the analysis of unexpected issues. This approach offers a deeper understanding of system performance and its internal functioning.

### 2.1. Monitoring vs Observability

Observability differs from monitoring. While monitoring focuses on collecting and analyzing predefined metrics to assess system health, it is often proactive, using alerts based on established thresholds [12]. For example, monitoring might trigger an alert to System Reliability Engineers when a server's CPU utilization exceeds 80%.

### 2.2. Importance of Observability

Traditional monitoring is insufficient for managing the complexity of modern systems. The advent of cloud computing and microservices architectures has significantly increased the complexity of distributed systems. Additionally, containerization technologies have introduced dynamic environments that scale to multiple instances on demand. Consequently, more advanced observability tools are required to manage these complex and dynamic distributed environments effectively. Enhanced observability allows for faster incident response, proactive issue detection, and quicker problem resolution.

### 2.2. Logging

Logs are discrete, timestamped records generated during a program's execution. They provide insights into system errors, environment details, and data related to interfacing systems, such as databases and API calls. Logs can be categorized into different levels, as outlined below.

Table 1. Log Levels [4]

| Log Level | Comments |
|-----------|----------|
| SEVERE | For fatal program errors |
| WARNING | For warning messages |
| INFO | For Informational messages during runtime |
| CONFIG | For Informational messages about config |
| FINE | Used for detailed info for debugging problems |
| FINER | Used for greater detailed info for debugging problems |
| FINEST | Used for greatest detailed info for debugging problems |

Logs enable System Reliability Engineers (SREs) and developers to examine specific events and trace what occurred at a particular moment, aiding in identifying the root cause of issues. To gather more detailed information, log levels may need to be adjusted. Modern applications allow for real-time adjustments to log levels without requiring a deployment or server restart. Various tools are available that aggregate logs from different systems, facilitating end-to-end debugging as requests pass through multiple systems and services in a distributed architecture.

### 2.3. Tracing

Traces capture the flow of a request as it moves through various systems and services in a distributed environment, including cloud-provided services. They offer a visual representation of how a request interacts with different services, databases, and external APIs, highlighting latency and failure points. Both open-source and paid tools are available that offer tracing capabilities. These tools typically require an agent to be installed within the application to collect the data and send it to a centralized server for analysis. [3].

### 2.4. Metrics

Systems or services generate numerical attributes known as metrics, which measure the performance and health of a system over time. These metrics provide a quantitative view of a system's performance, enabling the identification of potential issues. Alerts can be configured to trigger when the values of these metrics exceed specified thresholds.

### 2.5. Tools

Various tools are available for observability, both from cloud providers and third-party vendors. Cloud providers offer platform-specific tools, such as AWS CloudWatch, Google Cloud Monitoring and Azure Monitoring [9]. Apache SkyWalking [13] is a open source observability tool. Additionally, there are open-source tools like Prometheus. Paid options like New Relic and DataDog [9] offer more advanced features. Typically, data from these systems needs to be transformed and fed into a data analysis system for further study, although some paid tools also include advanced data analysis capabilities. In the dissertation, Moreira [10] evaluates various tools based on several factors, including usability, learning curve, integration, scalability, unified solutions, official documentation, community support, reliability, cost, maintenance support, implementation effort, integration, and visualization. The study concludes by recommending the use of OpenTelemetry due to its vendor-neutral nature and flexibility, allowing for easier transitions between vendors if the need arises.

## 3. SECURITY

Security practices safeguard enterprises from threats stemming from the growing reliance on technology. The following protection mechanisms are essential for ensuring a company's security:

### 3.1. Prevention against Cybercrimes

Cybercriminals exploit various methods to target enterprises, often leading to disruptions in services and financial losses. Security breaches can result in fines, lawsuits, and a loss of customer confidence, which can ultimately damage business relationships and partnerships.

### 3.2. Prevention against data breaches

Customer information, such as payment details, health records, and identity data, is frequently stored and processed online. Implementing robust security measures is crucial to prevent identity theft, fraud, and privacy breaches. Data breaches can severely damage an enterprise's reputation, eroding customer trust in the company. Additionally, they can lead to violations of legal and regulatory requirements such as GDPR (General Data Protection Regulation), HIPAA (Health Insurance Portability and Accountability Act), and PCI-DSS (Payment Card Industry Data Security Standard) [5], potentially resulting in significant penalties, fines, and legal actions.

## 4. SECURITY OBSERVABILITY

Security Observability involves extending observability practices to enhance system security by leveraging logs, metrics, and trace data. This approach may require collecting additional data to detect, investigate, and respond to security threats in real-time. It goes beyond traditional security monitoring by providing deeper insights into system behavior, enabling more proactive threat identification and response.

**4.1. Logging in the context of Security** Logs are valuable for detecting suspicious activities such as authentication attempts and unusual network behavior. They can provide a detailed history of user activities, system changes, and interactions, aiding in monitoring and security analysis.

### 4.2. Tracing in the context of Security

Tracing helps identify the path of a suspicious request and trace back the origin of the compromise. It can also be used to detect potential data leaks from the application by monitoring and tracing outgoing network calls.

### 4.3. Metrics in the context of Security

Metrics generated by analyzing various system parameters can help detect potential attacks. Indicators such as increased server CPU usage, multiple failed login attempts, or a spike in API error rates may signal an attack. Alerts triggered by these events, combined with trace data to identify the source, can help mitigate the attack effectively.

### 4.4. Sample Use Case

Figure 1 below depicts a sample application architecture consisting of multiple layers, starting with the web/app layer, followed by a Web Application Firewall, API Gateway, cloud infrastructure, cloud services, and applications hosted within the cloud environment.

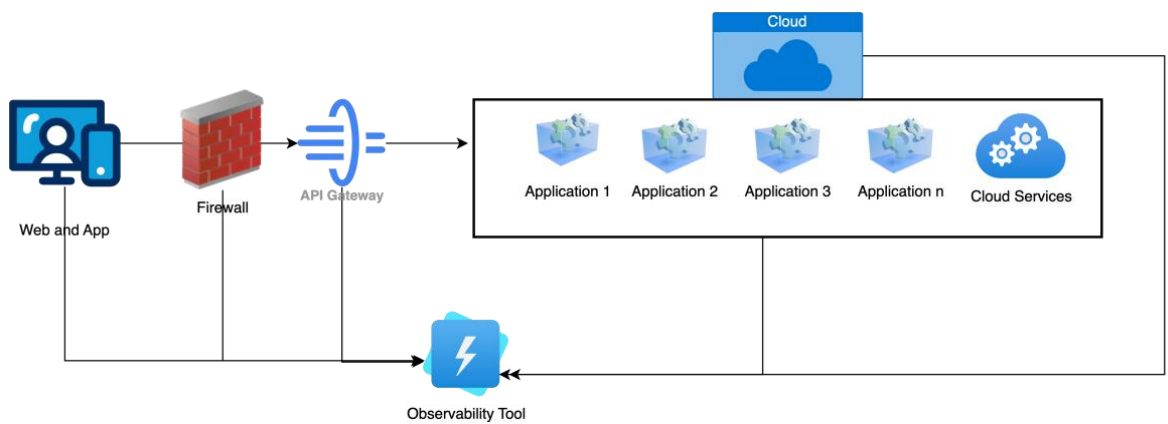

Figure 1. Sample Application

#### 4.4.1. Web and App

The Web and App layer represents the website and mobile applications, which are commonly built using open-source tools and technologies in most enterprises. Observability tools can collect logs generated by this layer and trace outgoing calls to track data exchanges originating from it. Reports, such as crash reports, can be generated to identify faulty SDKs within the app using these observability tools. The data collected through this process not only helps certify the reliability of the app or website but also enables monitoring for any suspicious activity.

### 4.4.2. Firewall

A Web Application Firewall (WAF) filters all incoming traffic from external sources, which can include both the internet and intranet. More advanced security configurations are necessary for traffic originating from the internet. Logs, traces, and metrics from the firewall can help identify patterns indicative of attacks, allowing the system to block the threat at the firewall level before it impacts the internal system, thus minimizing damage at the source.

### 4.4.3. API Gateway

The API Gateway intercepts all incoming API requests and forwards them to backend applications, which may be hosted either in the cloud or on-premise. Logs, metrics, and traces from the Gateway can help detect unusual patterns, such as an API receiving an excessive number of requests, which may indicate an ongoing attack. Telemetry data, including IP addresses and user agents, can be further analyzed to understand the nature of the attack. This data can also help detect potential data leaks, identifying whether unauthorized users are accessing sensitive information and enabling the blocking of such access. Security measures like rate limiting can reduce the impact of such attacks. Additionally, metrics like the number of 4XX or 5XX [6][12] errors can be used to monitor error rates, helping trace issues back to their source—whether it's a system problem or a brute force attack.

### 4.4.3. Application and Cloud Services

Applications hosted in the infrastructure can be either custom-built or third-party, while cloud services refer to the utilities provided by a cloud provider. The logs, metrics, and traces generated by these systems can help identify patterns related to security breaches. This data can reveal what information is exposed or compromised and detect users attempting to generate bot-like traffic. By analyzing these insights, actions can be taken to block such activities at earlier layers, before they reach the application layer.

### 4.4.4. Cloud

Cloud infrastructure, as well as on-premise infrastructure, generates logs, metrics, and traces that can be used to identify accessed applications and utilized cloud services. These logs can help trace network packets and monitor data movement, enabling the detection of data breaches, unauthorized access, or intrusions within the network.

### 4.4.5. Sample Metrics

Table 2 below presents common metrics used across systems to monitor system health and detect anomalies.

Table 2. Metrics

| Metrics Name | Description | System |
| --- | --- | --- |
| CPU Usage | CPU Usage percentage | Cloud Servers |
| Memory consumption | Memory Usage | Cloud Servers and Services |
| Request Latency | Latency in processing the request | Cloud Services and Application |
| Error Rates | Errors while processing the requests | API Gateway, Cloud Services and Application |
| Throughput | Number of requests processed in a minute | API Gateway, Cloud Services and Application |
| Network Packet IN/OUT | Number of Network packets received and sent by the system | Firewall Gateway, Cloud Services. |

## 5. USE CASE

Malin [8] discusses in the thesis how observability data can be highly effective in detecting attack patterns in IoT environments. In an experiment, observability data was fed into a Deep Neural Network model trained on attack data, resulting in a 92% accuracy rate for detecting attacks, compared to 50% accuracy using non-observability data. Similarly, Nousiainen [11] explains how observability in the cloud environment can support cloud governance, particularly for security auditing. A service called Security Hub collects logs, metrics, and traces from various systems. System analyses the data and generates events based on priority, sending them to SNS topics. In the cloud environment, configuration updates are also continuously monitored.

## 6. APPLICATION OF ML AND AI

The metrics, trace, and log data collected can be further analyzed using machine learning models for anomaly detection, time series forecasting, failure prediction, and security modeling [14]. According to a survey conducted by Ankur [14], these advancements have led to significant empirical improvements, with companies reporting incident detection rates as high as 60-70% and a reduction in user-impacting failures by 25-50%. These models can also be enhanced to detect security anomalies. Kinyua et al. [15] analyzed studies on the role of AI/ML in cybersecurity analytics and threat intelligence, highlighting how Artificial Neural Networks and Reinforcement Learning can be effectively utilized to detect security anomalies.

## 7. CHALLENGES

As applications grow, their architecture becomes increasingly complex, leading to a lack of understanding of the overall system and challenges in generating meaningful observability data for security purposes [12]. The application will produce vast amounts of data, which the observability system must transform and analyze. This analysis demands more sophisticated systems, and using machine learning models to achieve high accuracy with minimal false positives requires significant effort, infrastructure investments and trained resources.

## 8. FUTURE SCOPE

Security observability should be built into the practice of monitoring systems. Current observability tools need to be improved to spot security risks using advanced methods. Machine learning and artificial intelligence can help find patterns and stop some attacks at the first layer of defense.

## 9. CONCLUSIONS

As security incidents continue to rise, it is crucial for enterprises to invest in strengthening the security of both applications and infrastructure. Given the increasing complexity of systems, it is essential to implement multiple security mechanisms to address security breaches and traffic attacks in real time. In this context, Security Observability serves two key purposes: it monitors the overall health of systems while simultaneously using captured data to detect potential security breaches or leaks. Integrating Security Observability alongside traditional observability enhances system resilience and security.

# REFERENCES

[1] K. Thakur, M. Qiu, K. Gai and M. L. Ali, "An Investigation on Cyber Security Threats and Security Models," 2015 IEEE 2nd International Conference on Cyber Security and Cloud Computing, New York, NY, USA, 2015, pp. 307-311, doi: 10.1109/CSCloud.2015.71.

[2] Jouini, M., Rabai, L. B., & Aissa, A. B. (2014). Classification of Security Threats in Information Systems. *Procedia Computer Science*, *32*, 489–496. https://doi.org/10.1016/j.procs.2014.05.452

[3] Creane, B., & Gupta, A. (2021). *Kubernetes Security and Observability: A holistic approach to securing containers and cloud native applications*. O'Reilly Media, Inc.

[4] ONJava.com: An introduction to the java logging api. (n.d.). https://www.inf.ed.ac.uk/teaching/courses/ec/miniatures/logging-up.pdf, pp. 3-4

[5] Sargiotis, D. (2024). Legal and Regulatory Considerations in Data Governance. In *Data Governance: A Guide* (pp. 445–466). doi:10.1007/978-3-031-67268-2_15

[6] Security testing of web apis. (n.d.-b). https://repositorio-aberto.up.pt/bitstream/10216/153235/2/645851.1.pdf

[7] Defining an appropriate trade-off to overcome the challenges and limitations in software security testing. (2020). *Journal of Xidian University*, *14*(7). https://doi.org/10.37896/jxu14.7/166

[8] Strand, M. (2023). Observability in Machine Learning based Intrusion Detection Systems for RPL-based IoT.

[9] Mohamed, H. (2022). Towards an efficient multi-cloud observability framework of containerized microservices in kubernetes platform.

[10] Moreira, A. C. A. (2023). An observability approach for microservices architectures based on opentelemetry (Doctoral dissertation).

[11] Nousiainen, M. (2020). Improving Cloud Governance by Increasing Observability.

[12] Usman, M., Ferlin, S., Brunstrom, A., & Taheri, J. (2022). A survey on observability of distributed edge & container-based microservices. IEEE Access, 10, 86904-86919.

[13] Apache SkyWalking. Accessed: Aug. 17, 2022. [Online]. Available: https://skywalking.apache.org/

[14] Mahida, A. (2023). Machine Learning for Predictive Observability-A Study Paper. *Journal of Artificial Intelligence & Cloud Computing. SRC/JAICC-252. DOI: doi. org/10.47363/JAICC/2023 (2)*, *235*, 2-3.

[15] Kinyua, J., & Awuah, L. (2021). AI/ML in Security Orchestration, Automation and Response: Future Research Directions. *Intelligent Automation & Soft Computing*, *28*(2).

## Authors

Renjith Ramachandran received his Bachelor's Degree in Electronics and Communications Technology from India and his Master's Degree in Computer Science from the US. He spent 12 years as a consultant, taking on various roles from Software Engineer to Architect, and working with clients in industries such as Telecom, Banking, and Insurance. He currently serves as a Solutions Architect, with research interests that focus on software architectures, emerging technologies, and the development of innovative tools and frameworks.

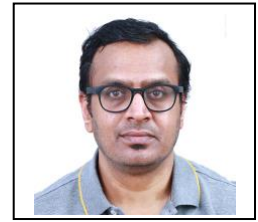